\documentclass[preprint,showpacs]{revtex4}
\usepackage{makeidx}
\usepackage{amsmath}
\usepackage{graphicx}
\usepackage{dcolumn}
\usepackage{bm}

\begin{document}

\title{Cosmological Bang within Matter Era. \\
Is the Generation of Galactic-Scale Mass Possible? }
\author{A. K. Avetissian}
\email{aavetis@ysu.am}
\affiliation{Department of Astrophysics, Yerevan State University, 1 A. Manoogian,
Yerevan 375025, Armenia}

\begin{abstract}
A heuristic hypothesis about domination of Bose-Einstein statistics in the
early Universe is suggested. The possibility of Bose-Einstein condensation
(BEC) of primordial baryon-antibaryon pairs is considered. In accordance
with this postulation enormous masses in the order of galactic mass may be
accumulated within the cosmic scales. At the certain threshold value of the
matter density the structural bosons decay into fermions and the sharp
breakdown of quantum-mechanical symmetry of the particles wave functions
occurs. Then, due to the Pauli principle of exclusion a large-scale phase
transition occurs because of enormous pressure jump of the matter. This
phenomenon might cause Cosmological Bang at the beginning stage of the
Matter Era.

As a mechanism of accumulation of galactic mass much larger than the
configuration with structural bosons, a hypothetical BEC of elementary
bosons (gauge bosons $W^{\pm }$\ and $Z^{0})$ is discussed as well.
\end{abstract}

\pacs{03.75.Hh, 14.70.-e, 98.80.Bp, 98.80.-k}
\maketitle



\section{Introduction}

As a possible version for observable evolution of the Universe an assumption
may be put forward that the Universe has been started-up from the Radiation
Era which probably may be discussed and commented in general arising from
the energetic convenience of Bose-Einstein statistics. Based on this
assumption, we predict that the evolution of the Universe has been started
with the \textquotedblleft Era of Bose-Einstein statistics\textquotedblright
. At the further period of Matter Era the Fermi-Dirac statistics began act,
and since this period the \textquotedblleft Era of Fermi-Dirac
statistics\textquotedblright\ dominates. In the result, like to Big Bang
phenomenon (because of abrupt breakdown of supersymmetry (SUSY) at the
earliest stage of Universe), an alternative Cosmological Bang at the
beginning stage of Matter Era might occur as a consequence of spasmodically
breakdown of quantum statistics. The great likelihood of Bose-Einstein
statistics is directly connected with the energetic convenience of the
particles population in the quantum microstates with possible lowest energy
without any restriction, in contrast to Fermi-Dirac statistics where the
Pauli principle excludes the occupation of the same energetic state by other
fermions. This is a well-known macro-scale quantum phenomenon of \
Bose-Einstein condensation (BEC).

The stable baryonic configurations for considering problem were usually
considered in assumption arising from the current state of the
matter--already formed steady-state nucleosynthesis and existing nuclei
within the model of Fermi-Dirac statistics. Such approach to the main
cosmological problem, as the generation of cosmic objects is, seems to be a
particular case of more general situation in the early Universe when before
the thermodynamic equilibrium between the radiation and matter the
Bose-Einstein and Fermi-Dirac statistics of primordial particles have been
disjointed because of enormous values of corresponding thermodynamic
potentials. This fact indicates in favor of initial large-scale phase
transition within the evolution process of early Universe. On the base of
above mentioned assumption concerning the Bose and Fermi statistics, and
taking into account that the population of the same quantum microstates by
the bosons stipulates incomparable less internal energy, a heuristic
hypothesis about dominating of Bose-statistics in the early Universe is
predicted. Such a physical assumption is in the accordance with the basic
process of production of the structural Bose-pairs of primordial particles
and antiparticles from the initial high energetic $\gamma $-photons in the
early Universe. The astrophysical conditions of the Universe at the
beginning stage of Matter Era satisfy the physical requirements for BEC of
such pairs.

For the beginning let \textrm{us} leave these heuristic abstractions
(detailed investigation will be done separately) and concretize on the
possibility of generation of structural bosons by the high energetic
primordial photons. The theoretical aspects and mechanisms developed for
electron-positron pairs production from high energy photons in the
superdense nuclear matter and multiphoton production by the nonlinear
channels in the strong radiation field of incomparable low frequencies than
the threshold one in diverse astrophysical cases (\cite{1,2,3,4};
H.K.Avetissian, A.K.Avetissian, et. all.) principally can be generalized for
the proton-antiproton pairs production. As to neutrons and antineutrons
pairs, it is obvious that the mechanisms of their production in principle
may be investigated through the other intermediate physical channels,
because they do not involve in the electromagnetic interaction (\cite{5};
W.Fauler, F.Hoyle; \cite{6}; G.Feinberg, L.Lederman). Since the high
energetic $\gamma $-photons may generate baryon-antibaryon pairs, and the
photons are electrically pure neutral, they could create just pure neutral
elementary particles--bosons, or even structural bosons with the total zero
electrical charge within appropriate elementary phase space.

\section{Theoretical prerequisites}

\subsection{BEC of structural bosons}

For the first quasi-qualitative/quasi-quantitative investigation we will
start from the case of structural bosons, consisted from proton, neutron,
electron and their antiparticles, respectively.

The spatial scale of quantum-electrodynamics phenomena, especially for $%
\gamma +\gamma \rightarrow p+\widetilde{p}$ process, has the same order
value as the proton's Compton wavelength (the average phase-space
localization of $\gamma $-photons): $\lambda _{p}=\hbar /m_{p}c\approx
2.1\times 10^{-14}\ \mathrm{cm}$. The physical approach about
\textquotedblleft point-or-elementary\textquotedblright\ assumption of
structural bosons will maintain descriptive strength until the scattering
amplitude of these structural bosons will exceed their own size by several
times. If the average size of baryon-antibaryon pairs is nearly $2\times
10^{-14}\ \mathrm{cm}$, the average distance between pairs at least must be
in order of $d_{p\widetilde{p}}\approx 5\times 10^{-14}\ \mathrm{cm}$, which
corresponds to the upper limit of pairs concentration $n_{p\widetilde{p}%
}\approx 1.9\times 10^{39}\ \mathrm{cm}^{-3}$.

For mentioned densities and supposed temperatures of the Universe $T\sim
10^{10}\div 2\times 10^{9}\ \mathrm{K}^{0}$\ at the beginning stage of
Matter Era (see \cite{7} Ya.Zel'dovich, I.Novikov; and \cite{8,9}
S.Weinberg) the baryon-antibaryon superdense degenerate plasma moderately
reaches its relativistic bound. These temperatures are relatively close to
the values $T_{\max }\sim 5\times 10^{8}\ \mathrm{K}^{0}$, obtained
theoretically in \cite{10} (V.Urpin, D.Yakovlev) and in \cite{11} (D.
Sedrakyan, A. Avetissian). Hence, the temperature of BEC may be obtained
based on generalized relativistic formula%
\begin{equation}
\frac{N}{V}=\frac{gT^{3}}{2\pi ^{2}\left( \hbar c\right) ^{3}}%
\int_{0}^{\infty }\frac{z^{2}dz}{e^{z}-1}=\frac{gT^{3}}{2\pi ^{2}\left(
\hbar c\right) ^{3}}\Gamma \left( 3\right) \zeta \left( 3\right) .  \label{1}
\end{equation}%
Here $c$\ is the velocity of light, $\hbar $\ is the Plank constant, $g=2s+1$%
- spin degeneracy factor, $\Gamma \left( x\right) $- Gamma function, and $%
\zeta \left( 3\right) $ $=1.202$ - Ryman's Zeta function (hereafter the
Boltzman's constant $k_{B}\equiv 1$). The temperature of BEC, obtained from (%
\ref{1}), is:%
\begin{equation}
T_{0}=\left( \frac{2\pi ^{2}}{2.4g}\right) ^{1/3}\hbar c\left( \frac{N}{V}%
\right) ^{1/3}=\left\{ 
\begin{array}{c}
4.63 \\ 
3.21%
\end{array}%
\right\} \times 10^{12}\ \mathrm{K}^{0}.  \label{2}
\end{equation}%
The number on upper row of (\ref{2}) corresponds to the spin-singlet state
of baryonic pairs, on lower row -- to the spin-triplet state. The relatively
higher temperature of BEC indicates in favor of energetic efficiency of
theoretical model, so the realization of spin singlet state of baryonic
pairs is more likelihood.

The spherical-symmetric configuration of baryon-antibaryon pairs (\textit{so
called hypothetical Universe}) may be qualitatively described in presumption
of convective stability; then the equations of thermodynamic and
hydrodynamic equilibrium without relativistic corrections get the following
form \cite{8}: 
\begin{subequations}
\label{3}
\begin{equation}
\frac{dm}{dr}=\frac{4\pi }{c^{2}}r^{2}\rho _{\varepsilon }\left( r\right) ,
\label{3.1}
\end{equation}%
\begin{equation}
\frac{dp}{dr}=-\frac{G}{c^{2}}\frac{1}{r^{2}}m\left( r\right) \rho
_{\varepsilon }\left( r\right) .  \label{3.2}
\end{equation}

Here $G$ is the Gravitation constant, $\rho _{\varepsilon }\left( r\right) $%
--BEC energy density of baryonic matter, $m\left( r\right) $--mass of the
matter within the central sphere with radius $r$. The system (\ref{3}) must
be also completed by the equation of Bose-Einstein condensate state with
baryon-antibaryon pairs: 
\end{subequations}
\begin{equation}
p_{_{BEC}}=\frac{gT^{4}}{6\pi ^{2}\left( \hbar c\right) ^{3}}\Gamma \left(
4\right) \zeta \left( 4\right) =\frac{g\pi ^{2}T^{4}}{90\left( \hbar
c\right) ^{3}}.  \label{4}
\end{equation}

It is easy to show that the photonic pressure is negligible in comparison
with the BEC one. Then it is essential to investigate the physical approach
when BEC of baryon-antibaryon pairs can be considered as an ideal gas. The
critical value of concentration when the amplitude of the pairs scattering
is of the order of their own size, corresponds to the Compton wavelength:

\begin{equation}
n_{cr}=\left( \frac{N}{V}\right) _{cr}\sim \frac{1}{\left( 4\pi /3\right)
\lambda _{p}^{3}}\sim 2.6\times 10^{40}\ \mathrm{cm}^{-3}.  \label{5}
\end{equation}

Up to this critical density the baryon-antibaryon pairs do not
\textquotedblleft sense\textquotedblright\ the own inner structure and may
be considered as elementary particles, so the equation of state of baryonic
matter (\ref{4}) can be presented in more generalized form:%
\begin{equation}
p_{_{BEC}}\left( r\right) =\frac{g\pi ^{2}T^{4}\left( r\right) }{90\left(
\hbar c\right) ^{3}}  \label{6}
\end{equation}%
The weak\textrm{\ }dependence on the radius of the functions $p\left(
r\right) $ or $T\left( r\right) $ evidences that within the BEC state
approximation the matter responds extremely weak on external pressure and in
the result the gravitational pressing may accumulate huge masses within the
cosmic scales.

The gravitational pressing of baryonic matter within the BEC and
accumulation of mass up to above mentioned critical densities may be
investigated within two alternative physical assumptions:

\textbf{1)} The process of gravitational pressing is assumed to be
isothermal,

\textbf{2)} The gravitational pressing varies the temperature of baryonic
matter from its initial value $T\sim 10^{10}\ \mathrm{K}^{0}$ up to final $%
T\sim 4.63\times 10^{12}\ \mathrm{K}^{0}$ (i.e. until the temperature of the
BEC).

Within the approximation of the \textbf{process 1}, at the end of
gravitational pressing (just before the quantum symmetry breakdown) the
pressure of baryonic matter in accordance with (\ref{6}) achieves its
extreme value

\begin{equation}
p_{_{BEC}}^{^{T=\mathrm{const}}}\approx 1.26\times 10^{24}\ \mathrm{Pa.}
\label{7}
\end{equation}%
In this intermediate state, such hypothetical configuration\ of
baryon-antibaryon degenerate plasma appears within extremely non-stable
state. Actually, because of scattering of pairs on each other, at the
critical value of density $n_{cr}\approx 2.6\times 10^{40}\ \mathrm{cm}^{-3}$
this state should fail due to the breakdown of quantum statistics of
Bose-pairs. This unstable state will be transferred from Bose-Einstein
statistics to Fermi-Dirac one and in the result the pressure of the fermions
system will increase significantly, in accordance with the Pauli exclusion
principle:

\begin{equation}
p_{_{F-D}}\approx 3.36\times 10^{35}\ \mathrm{Pa.}  \label{8}
\end{equation}%
From the formulas (\ref{7}) and (\ref{8}) we obtain the corresponding jump
of pressure in the phase transition within \textbf{process 1}, stipulated by
the phenomenon of quantum mechanical symmetry breakdown:

\begin{equation}
\frac{p_{_{F-D}}}{p_{_{BEC}}^{^{T=\mathrm{const}}}}\approx 2.27\times
10^{11}.  \label{9}
\end{equation}

\textbf{2)} At the end of gravitational collapse, the pressure achieves its
extreme value:

\begin{equation}
p_{_{BEC}}^{^{T\neq \mathrm{const}}}\approx 5.8\times 10^{34}\ \mathrm{Pa,}
\label{10}
\end{equation}%
so for the corresponding pressure jump from the formulas (\ref{10}) and (\ref%
{8}) we obtain

\begin{equation}
\frac{p_{_{F-D}}}{p_{_{BEC}}^{^{T\neq \mathrm{const}}}}\approx 5.8.
\label{11}
\end{equation}%
The comparison of expressions (\ref{9}) and (\ref{11}) shows that within the
presumed model of BEC the alternative cosmological Bang is realized
explicitly via quasi-isothermal \textbf{process 1}.

The results of numerical analyses of the theoretical model investigations
are represented in Fig. 1-2. Figure 1 represents the behavior of radial
pressure within the baryon-antibaryon configuration. Let specify: which
value of radius must interrupt the numerical integration of the system (\ref%
{3})? This question is equivalent to the physical issue: what value of $%
r_{\max }$ should be recognized as a \textquotedblleft effective radius of
baryon-antibaryon stable configuration\textquotedblright\ (i.e. $%
R_{_{conf}}^{^{eff}}$ $=r_{_{\max }}$)? Both issues depend on the minimal
value of pressure until which the approach of gaseous BEC may be considered
physically reliable yet. This condition is equivalent to the physical
requirement $T>>T_{_{cryst}}$, where $T_{_{cryst}}$ - is the temperature of
crystallization of the outer crust of baryonic configuration. Based on
analogues investigations (see e.g. \cite{12}) one can estimate for the upper
limit of this assumption $T_{_{cryst}}\leq 10^{8}\ \mathrm{K}^{0}$. In
accordance with the above mentioned, the minimal value of pressure obtained
from (\ref{4}): $p_{_{\min }}\leq 1.3\times 10^{16}\ \mathrm{Pa}$. Finally,
the criterion $p\geq p_{_{\min }}$ determines the lower limits both for the
effective radius ($R_{_{conf}}^{^{eff}}$) and effective mass ($%
M_{_{conf}}^{^{eff}}$ ) of the hypothetical Universe. As it is seen from
Fig. 1, for $p_{_{\min }}\leq 1.3\times 10^{16}\ \mathrm{Pa}$ the radius of
baryon-antibaryon configuration before the Cosmological Bang was
approximately $r_{_{\max }}\sim 10^{13}\mathrm{m}$.

Figure 2 represents the behavior of central mass of baryon-antibaryon
configuration within radius $r$. As it mentioned above, in accordance with
the condition $p\geq p_{\min }$ the integration of system (\ref{3}) may be
interrupted at $r_{\max }\sim 10^{13}\mathrm{m}$. Then the effective radius
of baryon-antibaryon stable configuration will be nearly about $%
R_{_{conf}}^{^{eff}}$ $\approx 10^{13}\mathrm{m}$, and the effective mass $%
M_{_{conf}}^{^{eff}}$ $\approx 10^{40}\mathrm{kg}$.

\subsection{Hypothetical BEC of elementary\textrm{\ }bosons}

Beside the structural bosons -- baryon-antibaryon pairs at the generation of
cosmic matter in the scope of proposed model one should probably take into
account the elementary\textrm{\ }bosons as well ( $W^{\pm }$ and\ $Z^{0}$).
As the gauge bosons $W^{\pm }$\ and $Z^{0}$\ are almost $100$\ times as
massive as the proton, so they may have significant contribution in the mass
accumulation. Although these bosons have a very short-life time, they may be
engaged in the phenomenon of BEC. These issues require an additional
theoretical investigation to constitute: might the elementary bosons become
stable particles at the extra high densities? The possibility of this exotic
phenomenon, i.e. production of elementary Bose-particles after the breakdown
of supersymmetry and separation of fundamental interactions requires an
additional discussion in following aspect: might the intermediate carriers
of weak interaction appear and be stabilized in superdense plasma?
Alternatively, might they play a significant role in the phenomenon of mass
generation and further accumulation of the matter until galactic masses? At
last, might the above-mentioned elementary bosons be responsible for the
phenomena of Dark Matter and Dark Energy?

For more substantial discussion of mentioned speculative ideas connected
with primordial role of the electroweak interaction in the evolution of the
Universe it is necessary to take into consideration the conditions at very
early stages of Universe. If about baryonic BEC one can speak at the
temperatures till $T\sim 10^{11}\ \mathrm{K}^{0}$, then above these
temperatures, namely, about $T\sim 10^{13}\ \mathrm{K}^{0}$ the physical
conditions allow the accumulation of sufficient number of $W^{\pm }$ and $%
Z^{0}$ bosons.

Very arrested alternative hypothesis may raise up regarding the mass
generation problem within celestial objects. For instance, might much more
densities of hypothetical plasma (compared with baryonic plasma!) consisted
mainly of $W^{\pm }$ and $Z^{0}$ bosons be a consequence of $100$ times
smaller (compared with baryons) value of their Compton-length $\lambda
_{w}=\hbar /m_{w}c\approx 2\times 10^{-16}\ \mathrm{cm}$ ?

In the same approximation of ideal nonrelativistic gas one can estimate the
lower bound for $W^{\pm }$\ and $Z^{0}$ bosons concentration and BEC
temperatures:%
\begin{equation*}
n_{\min }\sim 2\times 10^{45}\ \mathrm{cm}^{-3}\text{;\qquad }T_{0}\text{ }%
\sim \text{ }2.5\times 10^{14}\ \mathrm{K}^{0}.
\end{equation*}%
The corresponded to these values of densities and temperatures pressure of
such bosonic system appears to be%
\begin{equation*}
P(0)\text{ }\sim \text{ }3.6\times 10^{27}\ \mathrm{Pa}.
\end{equation*}%
Note that because of absence of experimental data about $W^{\pm }$,\ $Z^{0}$
bosons scattering cross sections for estimation of the upper bound of $%
W^{\pm }$,\ $Z^{0}$ bosons concentration and corresponding BEC temperatures,
the presented estimations are limited with the lower bound of bosons
concentration. Otherwise, the larger values of the latter might give us the
larger BEC temperatures and, consequently, would be physically more
realistic to take into consideration the BEC phenomenon in the more earlier
stages of Universe.

\begin{figure}[b]
\includegraphics{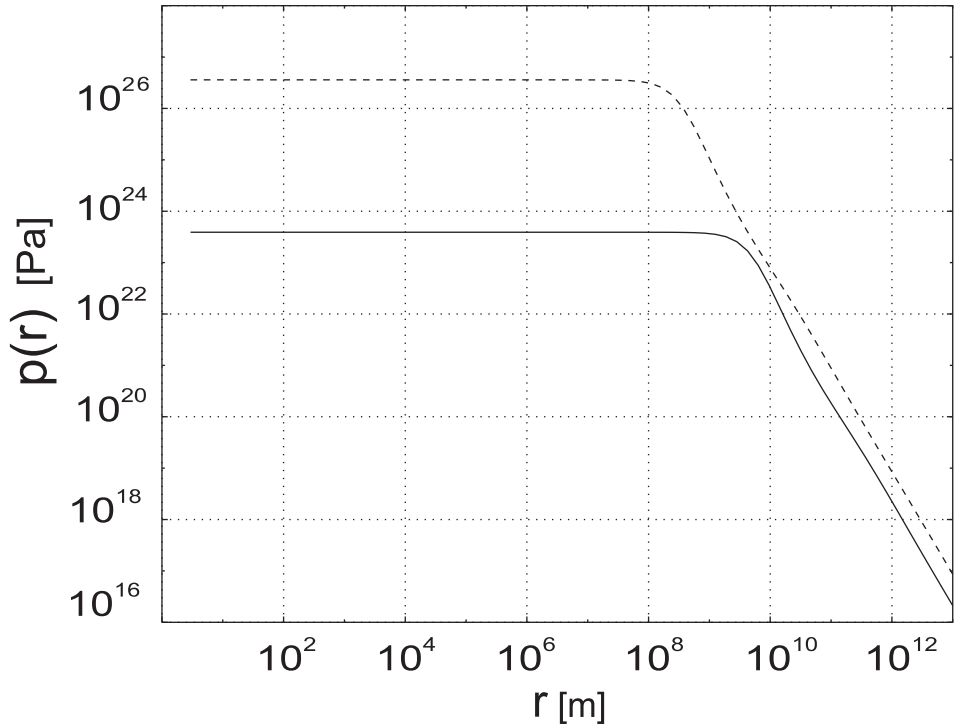}
\caption{The radial dependence of pressure for baryon-antibaryon (solid
line) and $W^{\pm }$,\ $Z^{0}$ bosonic (dashed line) configurations.}
\end{figure}

\begin{figure}[t]
\includegraphics{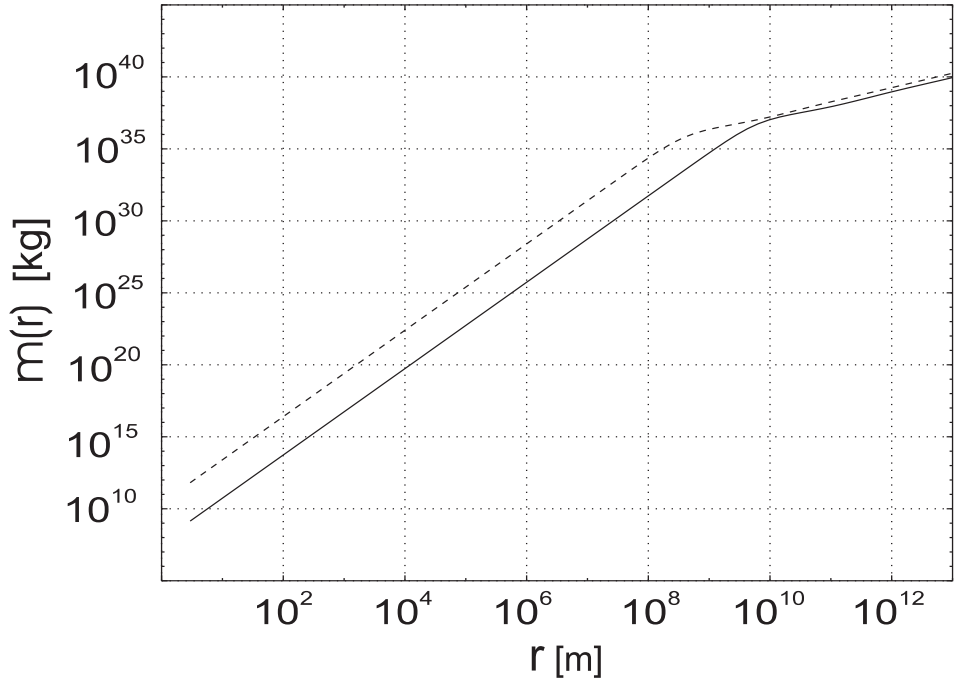}
\caption{The radial dependence of mass for baryon-antibaryon (solid line)
and $W^{\pm }$,\ $Z^{0}$ bosonic (dashed line) configurations.}
\end{figure}

The results of numerical treatments of BEC in large-scale cosmic matter (%
\textit{hypothetical Universe}) for both considered cases \textbf{A} and 
\textbf{B} are given in Fig. 1, 2. As is seen from the obtained graphics,
the accumulated mass due to the BEC of $W^{\pm }$,\ $Z^{0}$ bosons by two
order \ of magnitude exceeds the mass due to the baryonic BEC within the
radius $r$ $\sim $ $10^{10}\mathrm{m}$ of spherical symmetric configuration.

The stability of considered large-scale configuration will be a subject of
further investigation.

\section{Discussion}

It is physically clear that the possibility of generation of large-scale
configurations in principle does not except just in one local cosmic
space-time scales. Furthermore, this hypothesis and corresponding physical
model, as well as its astrophysical applications might have cosmological
consequences both within relatively small- and large-scales cosmic
structures. Note that after predicted Cosmological Bang at the beginning
stage of Matter Era the \textquotedblleft islands\textquotedblright\ of
similar configurations of degenerate matter might be accumulated, which
might form then various stable cosmic objects. It is physically obvious that
these objects principally might correlate in groups, clusters, or even
associations.

May the predicted Cosmological Bang be candidate for a \textquotedblleft
generator of galactic-scale masses\textquotedblright ? This physical
question requires additional analysis in the scope of considered hypothesis
and corresponding astrophysical model. Besides, the considered problems
require further investigation of the discussed analytical model and
development of the main ideas presented here in the scope of general
relativity. In addition, it is essential to take into consideration the
various empiric equations for BEC which might be more realistic from the
cosmological point of view.

\begin{acknowledgments}
This work was supported by International Science and Technology Center
(ISTC) Project No. A-1307.
\end{acknowledgments}

\end{document}